\documentclass[preprint,aps,showpacs,preprintnumbers,amsmath,amssymb,11pt]{revtex4}
\usepackage{fleqn,epsf}
\input{pstricks}
\usepackage{epsf}
\usepackage{natbib}%
\usepackage[dvips]{graphicx}
\begin{document}
\newcommand{\be}{\begin{equation}}
\newcommand{\ee}{\end{equation}}
\newcommand{\bea}{\begin{eqnarray}}
\newcommand{\eea}{\end{eqnarray}}
\newcommand{\nn}{\nonumber}
\newcommand{\dd}{\displaystyle}
\newcommand{\bra}[1]{\left\langle #1 \right|}
\newcommand{\ket}[1]{\left| #1 \right\rangle}
\newcommand{\qq}{<0|{\bar q} q|0>}
\newcommand{\spur}[1]{\not\! #1 \,}
\newcommand{\ttbs}{\char'134}

\title{$B\to K\eta,K\eta^{\prime}$ Decays}\thanks{Talk given at QCD08, the 14th International QCD Conference,
  7-12th July 2008, Montpellier (France)}
\author{T. N. Pham}
\affiliation{Centre de Physique Th\'eorique,
CNRS, Ecole Polytechnique, 91128 Palaiseau Cedex, France}       
\date{\today}
\begin{abstract}
The nonet symmetry scheme seems to describe rather well the masses and
$\eta-\eta^{\prime}$ mixing angle of the ground state pseudo-scalar 
mesons and is thus expected to be also a good approximation for
the matrix elements of the pseudo-scalar density operators 
which play an important role in charmless two-body $B$ decays with 
$\eta$ or  $\eta^{\prime}$ in the final state.
In this talk, I would like to report on a recent work on the 
$B^{-}\to K^{-}\eta, K^{-}\eta^{\prime}$ decay using  nonet symmetry for
the matrix elements of  pseudo-scalar density operators. We find that 
the branching ratio $B\to PP$, with an $\eta$ meson in the final state 
agrees well with data, while those with an $\eta^{\prime}$ meson
 are underestimated by $20-30\%$. This could be considered as a 
more or less successful 
prediction for QCDF, considering the theoretical 
uncertainties involved. This could also indicate that an additional 
power-suppressed terms could bring the branching ratio close 
to experiment, as with the $B\to K^{*}\pi$ and $B\to K^{*}\eta$ decay for 
which the measured branching ratios are  much bigger than the QCDF 
predictions.
\end{abstract}
\maketitle
\section{Introduction}
The $B \to K\eta,K\eta^{\prime}$ decays have been 
analysed in  recent papers \cite{Beneke,Zhu1,Kou1} in QCD 
Factorization (QCDF), in perturbative QCD (pQCD) \cite{Kou2,Li} and 
in soft collinear effective theory (SCET) \cite{Zupan}. In QCDF 
the $B\to K\pi$ branching ratio could be 
understood  with a moderate contribution from 
annihilation terms \cite{QCDF1,QCDF2}. Similarly, without fine tuning,  the   
$B\to K\eta^{\prime}$ branching ratio is predicted to be larger 
than that of $B\to K\pi$ in qualitative agreement with experiment, but is 
still underestimated by $20-30 \%$ compared with the measured value.

Apart from the power-suppressed $O(1/m_{b})$ annihilation terms, the main
theoretical uncertainties are the $B\to \eta^{\prime}$ transition
form factor and   the pseudo-scalar
density matrix elements for $\eta^{\prime} $. Historically, there is an
approximate $SU(3)$ relation between the octet pseudo-scalar 
density \cite{Gell-Mann} but there is no known explicit expression for 
the singlet pseudo-scalar density in  the nonet symmetry scheme. In 
this talk I would like to discuss a recent work \cite{Pham0} in which 
we show that nonet symmetry for the quark mass term in  
$\eta- \eta^{\prime}$  implies nonet symmetry for  the pseudo-scalar 
density matrix elements. With the
nonet symmetry  expression for the pseudo-scalar density matrix elements 
in  $\eta- \eta^{\prime}$, we obtain in QCDF a 
$B\to K\eta^{\prime}$  branching ratio, though sufficiently large, is still    
below the measured value by $20-30 \%$, but a large 
$B\to \eta'$ form factor  or additional power-suppressed 
terms could bring the predicted value closer to experiment.
\section{Nonet symmetry  in the $\eta-\eta^{\prime}$ system} 
Since QCD interactions through the exchange of gluons are 
flavor-independent, the wave function  for the pseudo-scalar meson nonet  
is also expected to be flavor-independent in the limit of vanishing
 current quark mass. The quark mass term  is the leading 
term in the large $N_{c}$ expansion while higher order terms 
in the chiral Lagrangian \cite{Gasser} is $O(1/N_{c})$ and is 
thus suppressed in the large $N_{c}$ limit. This justifies
the nonet symmetry for the pseudo-scalar meson mass matrix, the 
off-diagonal quark mass term $<\eta_{0}|H_{\rm SB}|\eta_{8}>$ then gives an 
 $\eta-\eta^{\prime}$ mixing angle $\theta= -18^{\circ}$ in good 
agreement with the value determined from the two-photon decay width
of $\eta$ and $\eta^{\prime}$ as mentioned in \cite{Pham0}. However, from 
the Gell-Mann-Okubo (GMO) mass formula, we would have 
\be
m_{\eta}^{2} = m_{8}^{2} - {\tan \theta}^{2}\,(m_{\eta'}^{2}- m_{8}^{2})
\label{etamass}
\ee
which gives, for $\theta = -18^{\circ}$, $m_{\eta}= 483\,\rm MeV $, 
about $60\,\rm MeV $ below experiment. This indicates that chiral
logarithms and chiral Lagrangian higher order terms \cite{Gasser,Gerard} 
which are second order in $SU(3)$ breaking as the $({\sin \theta})^{2}$ term 
in Eq.(\ref{etamass}) could contribute to $m_{8}$ and shift $m_{\eta} $ 
upward by a similar amount  with the result that the $\eta$ mass 
is  very close to the GMO value and a large $\eta-\eta^{\prime}$
mixing angle is obtained, rather than the small value of $ -10^{\circ}$ 
given by the GMO formula for $m_{8} $. We now use the  nonet symmetry 
mass term to derive  the pseudo-scalar density matrix element for 
$\eta,\eta^{\prime}$ which allows a calculation of $B \to K\eta^{\prime}$
as shown in the following sections.

The usual way to derive the pseudo-scalar density matrix elements 
between the vacuum and the pseudo-scalar meson nonet is to consider 
 the matrix elements of the divergence of the  axial vector currents 
between the vacuum and  pseudo-scalar meson nonet. For $\pi, K$ meson,
we have \cite{Pham0}:
\bea
 f_{\pi}B_{0}(m_{u} + m_{d}) &=& (m_{u}+ m_{d})\langle 0|\bar{u}\,i \gamma_5
 d|u\bar{d}\rangle ,\nonumber\\
 f_{K}B_{0}(m_{u} + m_{s}) &=& (m_{u}+ m_{s})\langle 0|\bar{u}\,i \gamma_5 s|u\bar{s}\rangle . 
\label{pi+}
\eea
and for $\pi^{0} $
\be
\  f_{u}B_{0}(m_{u} + m_{d}) = (m_{u}+ m_{d})\langle
0|\bar{u}\,i \gamma_5 u|u\bar{u}\rangle .
\label{pi0}
\ee
with  the $\pi$ and $K$ meson masses 
the usual expressions in terms of $B_{0}$ and the current quark 
mass \cite{Gasser,Donoghue}. To first order in 
$SU(3)$ breaking quark mass term, the decay constants $f_{q} $
($q=u,d,s $) is (putting $f_{q\bar{q}}= f_{q}$), 
\bea
&& f_{\pi} = f_{u\bar{d}}\approx f_{u},  \quad    
 f_{K}= f_{u\bar{s}}= (1 + \epsilon)\,f_{u\bar{d}}, \quad  \nonumber \\
&& f_{s}= (1 + 2\,\epsilon)\,f_{u}\approx (1 + \epsilon)\,f_{K}.
\eea

 Consider now the divergence of the $I=0$ axial vector current:
\be
 A_{{\rm n}\,\mu}= (\bar{u}\,\gamma_{\mu}\gamma_{5}u +
\bar{d}\,\gamma_{\mu}\gamma_{5}d), \quad
\quad A_{{\rm s}\,\mu}= \bar{s}\,\gamma_{\mu}\gamma_{5}s. \quad 
\label{Ans}
\ee
we have:
\bea
&&\kern -0.2cm \partial A_{\rm n} = 2(m_{u}\bar{u} i \gamma_5 u + 
m_{d}\bar{d} i \gamma_5 d)  + 2\frac{\alpha_s}{4\pi} 
G\,\tilde{G}. \quad \label{dAn} \\
&&\kern -0.2cm \partial A_{\rm s}  = 2 m_s\bar{s} i \gamma_5 s + 
\frac{\alpha_s}{4\pi} G\,\tilde{G}. \quad 
\label{dAns}
\eea
 Taking the matrix elements of $\partial A_{\rm n} $ and
 $\partial A_{\rm s}$ between the vacuum and $\eta_{0,8}$, we obtain: 
\bea
\kern -0.5cm f_{u}\frac{1}{\sqrt{3}}(m_{0}^{2} + B_{0}\frac{2}{3}(m_{s} + 
2{\hat  m}))& =& f_{u}\frac{1}{\sqrt{3}}m_{0}^{2} - f_{u}\frac{1}{\sqrt{6}}B_{0}\frac{2\sqrt{2}}{3}({\hat m}
-m_{s})+2\frac{1}{\sqrt{3}}{\hat m}\langle 0|\bar{u}\, i\gamma_5
 u|u\bar{u}\rangle ,\label{uu0} \\
\kern -0.5cm f_{s}\frac{1}{\sqrt{3}}(m_{0}^{2} + B_{0}\frac{2}{3}(m_{s} + 2{\hat m}))
&=& f_{s}\frac{1}{\sqrt{3}}m_{0}^{2} -  f_{s}\frac{2}{\sqrt{6}}B_{0}\frac{2\sqrt{2}}{3}({\hat m}
-m_{s}) +2\frac{1}{\sqrt{3}}m_{s}\langle 0|\bar{s}\, i\gamma_5
 s|s\bar{s}\rangle. 
\label{ss0}
\eea
 Similarly, for $\eta_{8}$ :
\bea
 f_{u}\frac{1}{\sqrt{6}}B_{0}\frac{2}{3}(2m_{s} + 
{\hat  m})&=& -f_{u}\frac{1}{\sqrt{3}}B_{0}\frac{2\sqrt{2}}{3}({\hat m}
-m_{s})+ 2\frac{1}{\sqrt{6}}{\hat m}\langle 0|\bar{u}\, i\gamma_5
 u|u\bar{u}\rangle ,\label{uu8} \\
 -f_{s}\frac{2}{\sqrt{6}}B_{0}\frac{2}{3}(2m_{s} + {\hat m})
& =&  -f_{s}\frac{1}{\sqrt{3}}B_{0}\frac{2\sqrt{2}}{3}({\hat m}
-m_{s})- 2\frac{2}{\sqrt{6}}m_{s}\langle 0|\bar{s}\, i\gamma_5
 s|s\bar{s}\rangle .
\label{ss8}
\eea
In deriving the above expressions, we have used the nonet symmetry 
mass formula, i.e $ m_{8}^{2} = B_{0}\frac{2}{3}\,(2m_{s} + \hat{m}) $, 
$m_{0}^{2} = {\bar m}_{0}^{2} + B_{0}\frac{2}{3}(m_{s} + 2\hat{m}) $
and $m_{08}^{2} = B_{0}\frac{2}{3}\sqrt{2}(-m_{s} + \hat{m}) $ where
$\hat{m}=(m_{u}+ m_{d})/2 $ and ${\bar m}_{0} $ is the anomaly
contribution to $m_{0}$, the singlet  $\eta_{0}$ mass. The  second 
term on the r.h.s of Eqs. (\ref{uu0}-\ref{ss0}) and the first term 
on the r.h.s of Eqs. (\ref{uu8}-\ref{ss8}) are the pole terms due to 
$\eta-\eta^{\prime}$ mixing. In the limit  $m_{u}=m_{d}=0$, the l.h.s 
and  r.h.s of Eq. (\ref{uu8}) become $f_{u}\,m_{8}^{2}/\sqrt{6}$ in 
agreement with the divergence equation Eq. (\ref{dAn}). 
Because of cancellation between the pole contribution and other 
 quark mass terms, Eqs. (\ref{uu0}-\ref{ss0}) are reduced to
the simplified form of  Eq. (\ref{pi+}) or  Eq. (\ref{pi0}). The  
pseudo-scalar density matrix elements in $\eta_{0}$ are then given by:
\be
\langle 0|\bar{u}\,i\gamma_5 u|u\bar{u}\rangle = B_{0}f_{u}, \quad
\label{u0}
\langle 0|\bar{s}\,i\gamma_5 s|s\bar{s}\rangle = B_{0}f_{s}.
\label{s0}
\ee
The same expression for $\eta_{8}$ is obtained similarly from
Eqs. (\ref{uu8}-\ref{ss8}). 
Thus,  $\langle 0|\bar{u}\,i\gamma_5 d|\pi^{+}\rangle$,
$\langle 0|\bar{u}\,i\gamma_5 u|\pi^{0}\rangle$ and 
$\langle 0|\bar{u}\,i\gamma_5 s|K^{+}\rangle $, and the matrix element
$\langle 0|\bar{u}\,i\gamma_5 u|u\bar{u}\rangle $ and 
$ \langle 0|\bar{s}\,i\gamma_5 s|s\bar{s}\rangle$ in $\eta_{0,8}$
are, apart from the  decay constant $f_{q}$, essentially the same, 
given by the parameter  $B_{0}$ and are consistent with nonet symmetry.

Since experimentally, $m_{08}^{2}= -(0.81\pm
0.05)\,m_{K}^{2}$ is rather close to the nonet symmetry value of 
$m_{08}^{2}\simeq -0.90\,m_{K}^{2}$ \cite{Donoghue}, we expect nonet 
symmetry for the pseudo-scalar density matrix elements in 
$\eta -\eta^{\prime}$ would be valid to this accuracy. 
Since  $m_{8}^{2}$  gets about $15\%$ increase from
higher order terms $L_{4},L_{5}, L_{6}, L_{8}$ and chiral 
logarithms,  Eqs. (\ref{uu8}-\ref{ss8}) show that 
$\langle 0|\bar{s}\,i\gamma_5 s|s\bar{s}\rangle $ in $\eta$ will be 
increased by a similar amount. A possible  similar $15\%$ increase for 
$m_{0}^{2}$ would also increase $\langle 0|\bar{s}\,i\gamma_5
s|s\bar{s}\rangle $ in $\eta_{0}$ by  a similar amount and would be  
additional source of enhancement for the $B\to K\eta^{\prime}$ 
branching ratio.

\section{The  $B^{-}\to K^{-}(\eta,\eta^{\prime})$ and 
 $ B^{-}\to \pi^{-}(\eta,\eta^{\prime})$ decays }
The  $B \to M_1 M_2$~ decay amplitude in QCD Factorization(QCDF) is
 given by\cite{QCDF1,QCDF2}:
\be
 {\cal A}(B \rightarrow M_1 M_2)=
 \frac{G_F}{\sqrt{2}}\sum_{p=u,c}V_{pb}V^{*}_{ps}\times 
  \left( \sum_{i=1}^{10} a_i^p
   \langle M_1 M_2 \vert O_i \vert B \rangle_H + 
 \sum_{i}^{10} f_B f_{M_1}f_{M_2} b_i \right ),
\label{BMM}
\ee
where the QCD coefficients  $a_{i}^{p}$ contain  vertex corrections,
 penguin corrections, and  hard spectator scattering contributions.
The hadronic matrix elements $ \langle M_1 M_2 \vert O_i \vert B
\rangle_H $  of the tree and penguin operators $O_{i}$ are given 
by factorization model \cite{Zhu1,Ali} and $b_{i}$ are annihilation terms.
The values for $a_{i}^{p}$, $p=u,c$  and $b_{i}$ computed from 
the expressions in \cite{QCDF1,QCDF2} at the renormalization 
scale $\mu=m_{b}$ and  with $m_{b}=4.2\,\rm GeV$ are given in the published 
work \cite{Pham0}. We estimate the  CKM matrix element $V_{ub}$ and the
CKM angle $\gamma$ from the $(db)$ unitarity triangle \cite{CKM}:
\be
\vert V_{ub}\vert=  \frac{\vert V_{cb}V_{cd}^{*}\vert}{\vert V_{ud}^{*}\vert} \vert  \sin \beta 
\sqrt{1+\frac{\cos^2 \alpha}{\sin^2 \alpha}} .
\label{Vub}
\ee
With $\alpha=(99^{+13}_{-9})^{\circ}$ \cite{PDG} and 
$\vert V_{cb}\vert = (41.78\pm 0.30\pm 0.08)\times 10^{-3}$ \cite{Barberio},
we find
\be
\vert V_{ub}\vert = 3.60\times 10^{-3}.
\label{Vub1}
\ee
which is quite close to  the exclusive data \cite{Barberio} 
$\vert V_{ub}\vert = (3.33-3.51)\times 10^{-3} $ .

Similarly, we  use the  current determination  $|V_{td}/V_{ts}|=
(0.208^{+0.008}_{-0.006})$ from the $B^{0}_{s}-\bar{B^{0}_{s}}$ mixing
measurements \cite{Abulencia} to obtain the  angle $\gamma$:
\be
\vert V_{td}\vert= \frac{\vert V_{cb}V_{cd}^{*}\vert}{\vert V_{tb}^{*}\vert} \vert  \sin \gamma
\sqrt{1+\frac{\cos^2 \alpha}{\sin^2 \alpha}}.
\label{Vtd}
\ee
which gives $\gamma = 66^{\circ} $ ( $|V_{tb}|=1 $) and $\alpha =
91.8^{\circ}$, in good agreement with the value found in the current
UT-fit value of $(88\pm 16)^{\circ}$. For other  parameters, we use
$m_{s}(\rm 2\,GeV)=80\,\rm MeV$,
$ f_{u}=f_{\pi}$, $f_{s}= f_{\pi}\left(1 +
  2(\frac{f_{K}}{f_{\pi}}-1)\right)$,
the current theoretical values \cite{Ball}:
\be
F^{B\pi}_{0}(0)= 0.258,\quad   F^{BK}_{0}(0) = 0.33, 
\label{FBK}
\ee
$F^{B\eta,B\eta^{\prime}}$   
from the $u$ quark content in $\eta$ and $\eta^{\prime}$: 
\be
F^{B\eta}(0)\kern -0.2cm = 0.58\,F^{B\pi}_{0}(0),F^{B\eta'}(0)\kern -0.1cm = \kern -0.1cm 0.40\,F^{B\pi}_{0}(0).
\label{FBeta}
\ee
and  the pseudo-scalar density matrix elements:
\be
\langle 0|\bar{s}\,i\gamma_5 s|\eta\rangle=\kern -0.1cm C_{\eta}B_{0}f_{s},
\langle 0|\bar{s}\,i\gamma_5 s|\eta^{\prime}\rangle=\kern -0.1cm C_{\eta^{\prime}}B_{0}f_{s}.
\label{O6}
\ee
obtained with  the $s$ quark content  $C_{\eta}=-0.57$, $C_{\eta^{\prime}}= 0.82$ 
and  an $\eta-\eta^{\prime}$  mixing angle $(-22\pm 3)^{\circ}$ \cite{Donoghue}
\begin{center}
\begin{table}[h]
\caption{ The branching ratio ${\cal B}(B\to P\eta, P\eta^{\prime})$ in QCDF}
\label{tab-res1}
\begin{tabular}{|c|c|c|}
\hline
Decay Modes &\kern -0.2cm QCDF BR ($\times 10^{-6}$)\kern -0.2cm
&\kern -0.3cm Exp. \cite{HFAG} \\
\hline
$B^{-}\to \pi^{-}\pi^{0}$ &$5.05$ & $5.7\pm 0.4$\\
$\bar{B}^{0}\to  K^{-}\pi^{+}$ &$18.25$ & $19.04 \pm 0.6$\\
$B^{-}\to \pi^{-}\eta$ &$3.39$ & $4.4\pm 0.4$\\
$B^{-}\to \pi^{-}\eta^{\prime}$ &$1.91$ & $2.6^{+0.6}_{-0.5}$\\
$B^{-}\to K^{-}\eta$ &$0.43$ & $2.2\pm 0.3$\\
$B^{-}\to K^{-}\eta^{\prime}$ &$48.26$ & $69.7^{+2.8}_{-2.7}$\\
\hline
\end{tabular}
\end{table}
\end{center}
As shown in Table \ref{tab-res1}, with a moderate annihilation 
term ($\rho_{A}=0.6$) and the current theoretical value for the 
$F^{B\pi} $ and $F^{BK} $ form factors \cite{Ball}, QCDF predictions 
are in reasonable agreement with experiment, except for the 
$B\to K\eta^{\prime}$ branching ratio which is underestimated 
by $20-30\%$. One could increase the 
$F^{B\eta^{\prime}} $ form factor to produce better agreement with 
experiment for $B^{-}\to \pi^{-}\eta^{\prime}$ for which the prediction
in Table \ref{tab-res1} is below the Babar value of 
$ (4.0 \pm 0.8\pm 0.4)\times 10^{-6} $ \cite{HFAG} and to bring the predicted 
$B\to K\eta^{\prime}$ branching ratio closer to experiment, but so far there
seems to be no evidence for a large $B\to \eta^{\prime}$ 
form factor compared with the nonet symmetry value as seen from the 
new Babar \cite{Babar}  upper limit 
${\cal B}(B^{+}\to \eta^{\prime}\ell^{+} \nu)/{\cal B}(B^{+}\to
\eta\ell^{+} \nu) < 0.57$ which is consistent with  nonet symmetry
for the $B\to \eta,\eta^{\prime}$ form factors given in Eq. (\ref{FBeta}).
\section{Conclusion}
We have shown that nonet symmetry for the
pseudo-scalar meson mass term implies nonet symmetry for the pseudo-scalar
density matrix elements. With this approximate relation, we
obtained  an improved estimate for  the $B\to P\eta^{\prime}$($P=K,\pi$ )
branching ratios. With  a moderate
annihilation contribution   consistent with the measured $B\to K\pi$ 
branching ratio, we find that a major part of the $B\to K\eta^{\prime}$
branching ratio could be obtained by QCDF. Without fine tuning 
or a large $F^{B\to \eta^{\prime}}$ form factor, we find that the 
$B\to K\eta^{\prime}$ branching ratio  is underestimated
by $20-30\%$. This could be considered as a more or less successful 
prediction for QCDF, considering the theoretical uncertainties involved.
This could also indicate that an additional 
power-suppressed terms could bring the branching ratio close 
to experiment, as with the $B\to K^{*}\pi$ \cite{Zhu3} and $B\to
K^{*}\eta$ decay for which the measured branching ratios are  
much bigger than the QCDF prediction.

\bigskip

\section{Acknowledgments}
I would like to thank  S. Narison and 
the  organizers of QCD 2008 for the warm hospitality extended to me 
at Montpellier. This work was supported in part by the EU 
contract No. MRTN-CT-2006-035482, "FLAVIAnet".

\end{document}